\documentclass[aps,prd,preprint,showpacs,floats,nofootinbib]{revtex4}
\usepackage{graphicx,subfigure}
\usepackage{bm}
\usepackage{euscript,amsmath}
\newcommand{\gsim}{\mathrel{\hbox{\rlap{\lower.55ex \hbox{$\sim$}} \kern-.3em \raise .4ex \hbox{$>$} } } }
\begin{document}
\title{{\LARGE A Model Independent Method to Study Dark Matter induced Leptons and Gamma rays }}

\date{\today}

\author{Mingxing Luo}

\author{Liucheng Wang}\email{liuchengwang@gmail.com}\thanks{(Corresponding Author)}

\author{Guohuai Zhu}

\affiliation{Zhejiang Institute of Modern Physics, Department of Physics, Zhejiang
University, Hangzhou, Zhejiang 310027, P.R.China}

\begin{abstract}
By using recent data, we directly determine the dark matter (DM) induced $e^\pm$ spectrum at the source from experimental measurements at the earth, without reference to specific particle physics models.
The DM induced gamma rays emitted via inverse Compton scattering are then obtained in a model independent way. 
However the results depend on the choice of the astrophysical $e^\pm$ background, which is not reliably known. 
Nevertheless, we calculate, as an illustration, the fluxes of gamma rays from the Fornax cluster in the decaying DM scenario with various astrophysical $e^\pm$ backgrounds. 
Without any assumptions on details of the DM model, the predictions turn out to be either in disagreement with or only marginally below the upper limits measured recently by the Fermi-LAT Collaboration. 
In addition, these DM induced ICS gamma rays in the GeV range are shown to be almost independent of choices of cosmic ray propagation model and of DM density profile, when a given astrophysical $e^\pm$ background is assumed. 
This provides a strong constraint on decaying DM scenario as the gamma rays may be produced in other processes besides inverse Compton scattering, such as the bremsstrahlung and neutral pion decays.
\end{abstract}
\maketitle

As one of the dominant components of the universe, dark matter (DM) has yet to show its existence other than its gravitational effects.
The nature of DM can be explored via searches at colliders, as well as via direct and indirect detection experiments.
Recently, indirect detection of DM has attracted great attention due to the cosmic ray electron/positron excesses observed
by the PAMELA \cite{Adriani:2008zr} and Fermi satellites \cite{Abdo:2009zk,Ackermann:2010ij}.
But the interpretation of these experimental results is subtle.
It is not easy to exclude the possibility that these excesses may origin from nearby astrophysical sources.
Even assuming the DM annihilation/decay to account for the PAMELA and Fermi-LAT observations, one has to face particle physics model dependence.

In this paper we will show that, it is possible to tightly constrain the decaying DM interpretation of electron/positron excesses in a particle physics model independent way,
by considering the gamma rays from nearby clusters.
In other words, constraints can be obtained without any assumptions on details of the DM model.

Experimentally, the PAMELA and Fermi-LAT Collaborations measure only the energy spectra of cosmic rays at the earth.
To compare with theoretical predictions,
one usually starts from a specific DM model to calculate the fluxes at the source and their propagation through the Galaxy. 
For a given astrophysical $e^\pm$ background, such a specific DM model should fit the observed $e^\pm$ spectrum at the earth.
Obviously, it is much desired to extract their fluxes at the source where they are generated in a model independent way. 
Actually, $e^\pm$ fluxes at the source can be obtained by solving an integral equation analytically, without introducing a specific DM model \cite{Hamaguchi:2009jb}\footnote{ 
We didn't notice the paper \cite{Hamaguchi:2009jb} until the first version of this paper appeared on arXiv. See the ``Note added'' for more details. }.
In this paper, we slightly improve this kind of method and apply it to updated experimental data.
Moreover, by taking $e^\pm$ fluxes at the source as an input, gamma rays emitted by these DM-induced energetic leptons via inverse Compton scattering (ICS) can be predicted independent of any DM model.
We show that the predictions of gamma rays turn out to be either in disagreement with or only marginally below the upper limits measured recently by the Fermi-LAT Collaboration \cite{Ackermann:2010qj}.
This DM-model independent method could be applicable to both annihilating and decaying DM scenarios, but we will focus only on decaying DM scenario in this paper. The discussion about annihilating DM case should be very similar to that of decaying DM.

Conventionally, the $e^\pm$ propagation in the Galaxy is governed approximately by the  diffusion equation
\begin{align}\label{eq:diffusion}
K(E)\cdot\nabla^{2}f_{e}^{\mathrm{DM}}(E,\vec{r})+\frac{\partial}{\partial E}\left[B(E)f_{e}^{\mathrm{DM}}(E,\vec{r})\right]+Q_{e}^{\mathrm{DM}}(E,\vec{r})=0~.
\end{align}
Here $f_{e}^{\mathrm{DM}}(E,\vec{r})$ is the DM-induced $e^\pm$ number density per unit energy. $K(E)$ stands for the diffusion coefficient,
which can be parameterized as $K(E)=K_{0}(E/\mathrm{GeV})^{\alpha}$ with $K_0$ and $\alpha$ given in Table \ref{tab:1}.
$B(E)$ describes the energy loss, which is effectively given as $B(E)=E^{2}/(\mathrm{GeV}\cdot\tau_{E})$, with $\tau_{E}=10^{16}$ s being a typical time scale in the Galaxy.
For decaying DM scenario, the source term $Q_{e}^{\mathrm{DM}}(E,\vec{r})$ can be expressed as
\begin{align}\label{eq:source}
Q_{e}^{\mathrm{DM}}(E,\vec{r})=\rho^{\mathrm{DM}}(\vec{r})\underset{i}{\sum}\frac{\Gamma_{i}^{\mathrm{DM}}}{M^{\mathrm{DM}}}
\frac{dN_{i}^{\mathrm{DM}}}{dE}=\rho^{\mathrm{DM}}(\vec{r})\, X(E)~.
\end{align}
Here $\rho^{\mathrm{DM}}(r)$, $\Gamma_{i}^{\mathrm{DM}}$, $M^{\mathrm{DM}}$
and $dN_{i}^{\mathrm{DM}}/dE$ are the DM density, the decay width
of a particular decay channel, DM particle mass and the $e^{\pm}$ spectrum
per DM decay via a particular channel, respectively. The summation
is over all possible decay channels and $X(E)$ contains all the particle physics information about DM.

\begin{table}
\begin{tabular}{|c|c|c|c|}
\hline
Model & $\alpha$ & $K_{0}$ in $\mathrm{kpc^{2}/Myr}$ & $L$ in $\mathrm{kpc}$\tabularnewline
\hline
\hline
MIN & 0.55 & 0.00595 & 1\tabularnewline
\hline
MED & 0.70 & 0.0112 & 4\tabularnewline
\hline
MAX & 0.46 & 0.0765 & 15\tabularnewline
\hline
\end{tabular}
\caption{\label{tab:1} Parameters in propagation models. MIN/MED/MAX refer to models which yield
minimal/medium/maximal positron flux, respectively \cite{Delahaye:2007fr}.}
\end{table}

Usually, $X(E)$ is determined by assuming a specific DM model. 
Then the DM induced $e^\pm$ at the earth can be determined by solving Eq. (\ref{eq:diffusion}) in a solid flat cylinder \cite{Delahaye:2007fr,Hisano:2005ec,Ishiwata:2009vx}
as\footnote{
In practice, one has to truncate the infinite series to a finite sum.
When $E' \simeq E$, the series in Eq. (\ref{eq:Solution}) converges very slowly since there is no exponential suppression.
In this range the solution is better expressed in an alternative form \cite{Ishiwata:2009vx}
$$
f_{e}^{'\mathrm{DM}}(E,\overrightarrow{r_{\odot}})=\frac{\tau_E}{E^2} \int_E^\infty dE' X(E') \exp\left[\frac{K_{0}\,\tau_{E}}{1-\alpha}\left(E^{\alpha-1}-(E')^{\alpha-1}\right)\nabla^{2}\right]\,
\left . \rho^{\mathrm{DM}}(\vec{r})\,\right \vert_{\vec{r}=\overrightarrow{r_{\odot}}}~.
$$
Taking the MED propagation model and Navarro-Frenk-White (NFW) DM density profile \cite{Navarro:1995iw} as an illustration,
and reordering the series in Eq. (\ref{eq:Solution}) from small to large $\vert \lambda_{mn} \vert$,
we shall take the first $1413$ terms of the series in Eq. (\ref{eq:Solution}) as a good approximation.
This truncated sum agrees well with $f_e^{'\mathrm{DM}}$ within $0.1\%$ error in the range $E' \simeq E$.}
\begin{align}\label{eq:Solution}
f_{e}^{\mathrm{DM}}(E,\overrightarrow{r_{\odot}})=\frac{\tau_{E}}{E^{2}}\overset{{\scriptscriptstyle \infty}}{\underset{{\scriptscriptstyle {\scriptstyle m,n=1}}}{\sum}}B_{mn}\int_{{\scriptscriptstyle E}}^{{\scriptscriptstyle \infty}}dE'\exp\left[\lambda_{mn}\left(E^{\alpha-1}-(E')^{\alpha-1}\right)\right]\, X(E')~,
\end{align}
where
\begin{align}\label{eq:Lamda}
B_{mn} & =  \frac{2 \sin (m\pi/2)}{J_{1}^{2}(\zeta_{n})R^{2}L}J_{0}\left(\frac{\zeta_{n}\, r_{\odot}}{R}\right) \int_{{\scriptscriptstyle 0}}^{{\scriptscriptstyle R}}dr\, r\int_{{\scriptscriptstyle -L}}^{{\scriptscriptstyle L}}dz\:\rho^{\mathrm{DM}}(\sqrt{r^{2}+z^{2}})J_{0}\left(\frac{\zeta_{n}\, r}{R}\right)\sin\left[\frac{m\pi}{2L}(z+L)\right]~, \nonumber \\
\lambda_{mn} & =  \left(\frac{\zeta_{n}^{2}}{R^{2}}+\frac{m^{2}\pi^{2}}{4L^{2}}\right)K_{0}\,\tau_{E}\,\frac{1}{\alpha-1},~
\end{align}
with the cylinder coordinates $z\in[-L,~L]$ in the $z$-direction and $r\in[0,~R]$ $(R=20~\mbox{kpc})$
in radius. Here $J_n$ is the $n$-th order Bessel function and $\zeta_{n}$'s are
successive zeros of $J_{0}$. The solar system is at $r_{\odot}=8.5~\mbox{kpc}$.

Surprisingly, the DM-induced $e^\pm$ spectrum $X(E)$ at the source can be determined in a DM-model independent way once $f_{e}^{\mathrm{DM}}(E,\overrightarrow{r_{\odot}})$ is known \cite{Hamaguchi:2009jb} .
Eq. (\ref{eq:Solution}) is actually the so-called Volterra integral equation and its inverse solution can be obtained analytically as
\begin{align}\label{eq:X}
X(E)=\frac{dg(E)}{dE}+(\alpha-1)E^{\alpha-2}\int_{\infty}^{E}dE'\frac{dg(E')}{dE'}R\left(E^{\alpha-1}-(E')^{\alpha-1}\right)~,
\end{align}
where\footnote{In practice, the infinite series will be truncated, in the same vein of Eq. (\ref{eq:Solution}).}
\begin{align}\label{eq:X1}
g(E)=-\frac{E^{2}}{\tau_{E}}f_{e}^{\mathrm{DM}}(E,\overrightarrow{r_{\odot}})\left / \sum \limits_{m,n=1}^\infty B_{mn} \right . ~, \hspace*{0.8cm}
R(x)=\mathrm{\boldsymbol{L}}^{-1}\left[\frac{1}{p\widetilde{K}(p)}-1\right]~,
\end{align}
with
\begin{align}\label{eq:X2}
\widetilde{K}(p)=\mathrm{\boldsymbol{L}}\left[K(x)\right]=\mathrm{\boldsymbol{L}}
\left[ \sum \limits_{m,n=1}^\infty B_{mn} \exp[\lambda_{mn}x] \left / \sum \limits_{m,n=1}^\infty B_{mn} \right . \right]~.
\end{align}
The source spectrum $X(E)$ can then be determined from the DM-induced $e^\pm$ at the earth with energies larger than $E$.
Here $\boldsymbol{L}$  denotes the Laplace transform and $\boldsymbol{L}^{-1}$ its inverse.
$\boldsymbol{L}$ can be performed trivially while the Cauchy's residue theorem is needed to perform $\boldsymbol{L}^{-1}$ analytically.
This part constitutes one of the major technical hurdles of our analysis. We refer to the Appendix for more details about the inverse solution of Volterra integral equation.

On the other hand, $f_{e}^{\mathrm{DM}}(E,\overrightarrow{r_{\odot}})$ can be obtained
by subtracting off the astrophysical $e^\pm$ background from the observed $e^\pm$ spectrum at the earth. The Fermi-LAT Collaboration have reported the $e^{\pm}$ spectrum in the range from $7$ GeV to $1$ TeV \cite{Abdo:2009zk,Ackermann:2010ij}. 
However our current understanding of the astrophysical $e^{\pm}$ backgrounds is still quite limited. 
As an illustration, we first take the conventional ``model 0'' \cite{Grasso:2009ma} of the $e^{\pm}$ background, which can be parameterized as \cite{Ibarra:2009dr}
\begin{align}
\Phi_{e^{-}}^{\mathrm{bkg}}(E) & =  \frac{82.0\epsilon^{-0.28}}{1+0.224\epsilon^{2.93}}~\\
\Phi_{e^{+}}^{\mathrm{bkg}}(E) & =  \frac{38.4\epsilon^{-4.78}}{1+0.0002\epsilon^{5.63}}+24.0\epsilon^{-3.41}\nonumber
\end{align}
in units of $\mbox{GeV}^{-1}\mbox{m}^{-2}\mbox{s}^{-1}\mbox{sr}^{-1}$ with $\epsilon=E/1\;\mbox{GeV}$.
The total $e^{\pm}$ background flux at the Earth can then be expressed as
\begin{align}
\Phi_{e^{\pm}}^{\oplus}(E_{\oplus})=\frac{E_{\oplus}^{2}}{E^{2}}
\left[\Phi_{e^{+}}^{\mathrm{bkg}}(E)+\mathrm{N}\times\Phi_{e^{-}}^{\mathrm{bkg}}(E)\right]
\end{align}
with a normalization factor $\mathrm{N}$.
To account for the solar modulation effects, the force field approximation $E_{\oplus}=E+e\,\phi_{F}$ with $\phi_{F}=0.55$ GV has been taken.
In order to leave room for the additional DM component below 100 GeV, we choose the normalization factor $\mathrm{N}=0.8$.
With this astrophysical $e^{\pm}$ background,
the introduction of an additional leptonic component from decaying DM could provide a plausible interpretation
of not only Fermi-LAT $e^\pm$ excess but also PAMELA anomaly in the positron fraction (See, e.g., \cite{Ibarra:2009dr,Luo:2009xd,Cheng:2010mw}).

\begin{figure}[tb]
\includegraphics[scale=1.0]{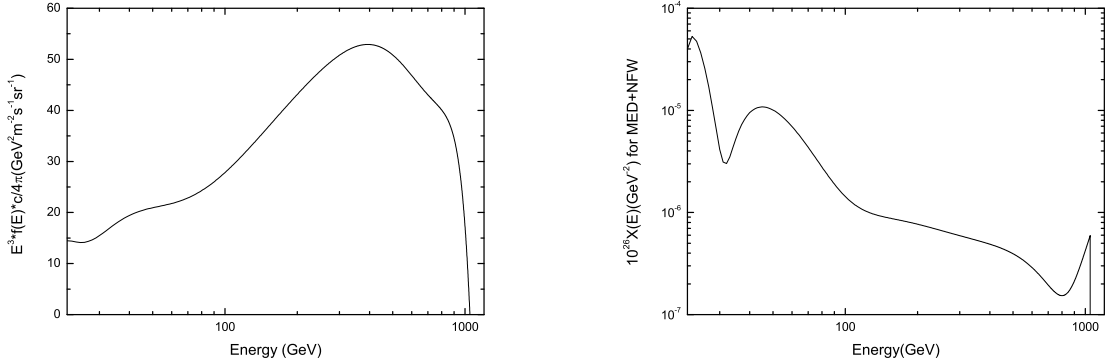}
\caption{\label{fig:fX} Left: $f_e^\mathrm{DM}(E,r_{\odot})$ extracted from the Fermi-LAT $e^\pm$ spectrum by subtracting off the background $\Phi_{e^{\pm}}^{\mathrm{bkg}}(E)$.
Right: $X(E)$ determined from $f_{e}^\mathrm{DM}(E,r_{\odot})$, assuming the MED propagation model and NFW DM density profile.}
\end{figure}

\begin{figure}[tb]
\includegraphics[scale=1.0]{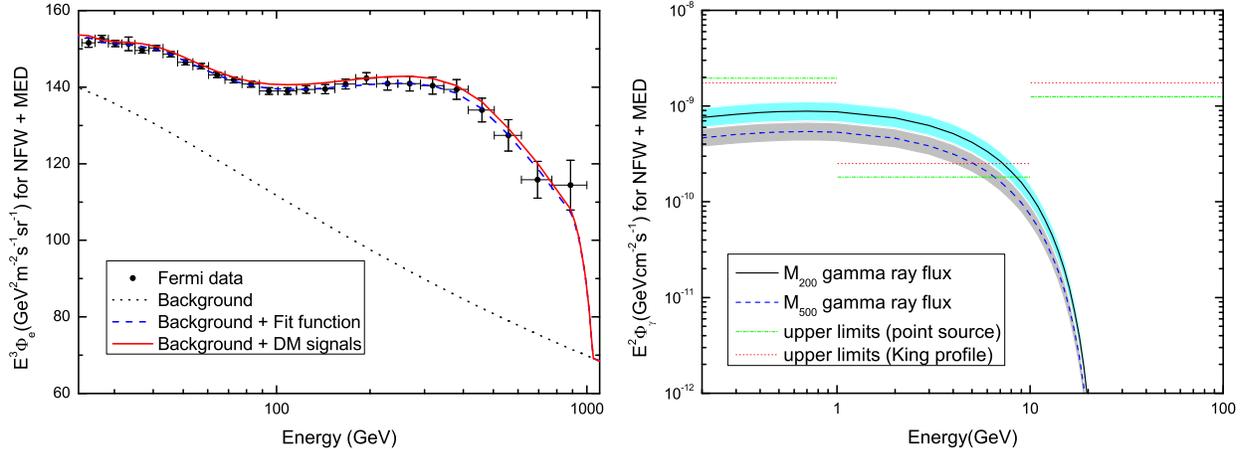}
\caption{\label{fig:NFW MED} NFW halo profile and MED propagation model are assumed. Left: The difference between the red solid
line and the blue dashed line can be viewed as a demonstration of the theoretical error in determining $X(E)$, as explained in the text.
Right: The correspondingly predicted ICS flux of photon is shown in the Fornax cluster. Experimental upper limits are taken from \cite{Ackermann:2010qj}. }
\end{figure}

Shown in the left part of Fig. \ref{fig:fX} is a fit function of $f_{e}^{\mathrm{DM}}(E,\overrightarrow{r_{\odot}})$
obtained by subtracting off $e^{\pm}$ background from the Fermi-LAT data.
Taking this fit function $f_{e}^{\mathrm{DM}}(E,\overrightarrow{r_{\odot}})$ as an input, one may obtain $X(E)$ via Eq. (\ref{eq:X}).
Shown in the right part of Fig. \ref{fig:fX} is the $X(E)$ thus obtained
for the MED propagation model and NFW DM density profile normalized with local DM density $\rho_\odot=0.3$ GeV/$\mbox{cm}^3$.
As discussed in the Appendix, we have made certain approximations in obtaining $X(E)$.
To estimate the theoretical errors, we have taken $X(E)$ as an input in Eq. (\ref{eq:Solution}) to get a new $f_{e}^{\mathrm{DM}}(E,\overrightarrow{r_{\odot}})$.
Shown in the left part of Fig. \ref{fig:NFW MED} is a comparison of this new $f_{e}^{\mathrm{DM}}(E,\overrightarrow{r_{\odot}})$ with the original fit function.
One sees clearly that the errors are very small, never beyond few percents.

Taking this spectrum function $X(E)$ as an input, the ICS gamma rays can be deduced from the scattering of energetic $e^\pm$ on starlight and CMB photons.
One can then check these predictions against experimental measurements of gamma rays from inside/outside the Galactic halo.
We remind that the constraints obtained in this way does not depend on any details of the DM model.
Recently, Fermi-LAT Collaboration has measured gamma rays from nearby clusters of galaxies with an 18-month data set \cite{Ackermann:2010qj}.
These clusters are supposed to be highly DM dominated and isolated at high galactic latitudes.
High signal-to-noise ratios are anticipated for gamma-ray observations targeting nearby clusters.
Recent model-dependent studies \cite{Dugger:2010ys,Ke:2011xw} have shown that gamma rays from the Fornax cluster provide the strongest constraint for decaying DM.
In the following we will focus on the DM induced gamma rays from the Fornax cluster.
Certainly, there may exist other sources in clusters that can emit gamma-rays, besides DM annihilation/decay.
Nevertheless the ICS gamma-rays predicted from $X(E)$ can give theoretical lower limits on the gamma ray flux.
In the Fornax cluster, the ICS gamma rays comes mainly from the scattering of $e^\pm$ on the CMB photons,
while the effects of dust and starlight can be neglected \cite{Ke:2011xw,Pinzke:2011ek}.
Treating the Fornax cluster as a point source, we follow the same method in \cite{Ke:2011xw} to calculate the ICS gamma rays semi-analytically.
Shown in the right part of Fig. \ref{fig:NFW MED} is the predicted gamma ray spectrum,
which seems to disagree with the Fermi-LAT measurements of gamma rays \cite{Ackermann:2010qj} in the range of $1-10$ GeV.
Here we have considered the uncertainties from the total DM mass of the Fornax cluster. The corresponding viral masses $M_{200}$, $M_{500}$ and their error bars are adopted from \cite{Reiprich:2001zv}.

\begin{figure}[tb]
\includegraphics[scale=1.0]{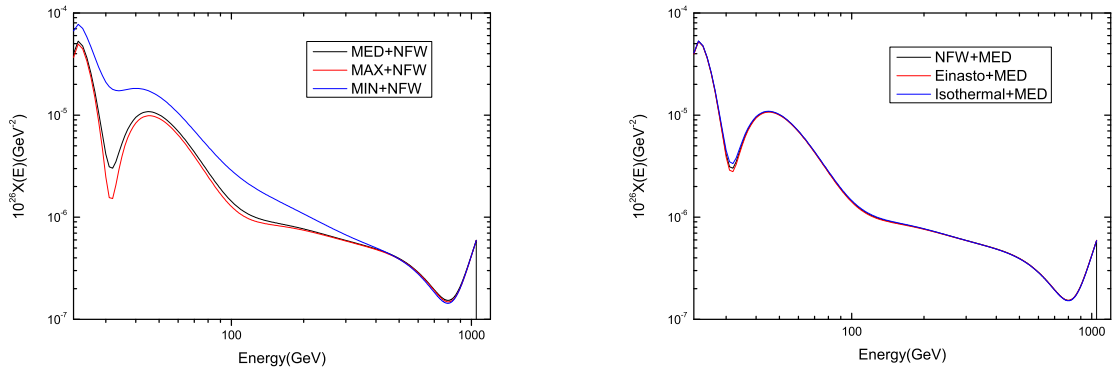}
\caption{\label{fig:error} Astrophysical uncertainties for the determination of $X(E)$ from $f_{e}^\mathrm{DM}(E,r_{\odot})$.
Left: NFW DM density profile is assumed while propagation models are varied.
Right: The MED propagation model is assumed while DM density profiles are varied.}
\end{figure}

We now address other relevant astrophysical uncertainties about the ICS of $e^\pm$ on the CMB from the Fornax cluster.
As the spectrum of CMB photons is well known, the main uncertainties arise from choices of propagation model and of DM halo profile in
determination of $X(E)$ from $f_{e}^{\mathrm{DM}}(E,\overrightarrow{r_{\odot}})$.
Shown in the left part of Fig. \ref{fig:error} are the $X(E)$'s obtained by using the MED, MIN and MAX propagation models respectively, with the default NFW DM density profile.
Shown in the right part of Fig. \ref{fig:error} are the $X(E)$'s corresponding to the NFW, Einasto \cite{Graham:2005xx} and
Isothermal \cite{Bahcall:1980fb} density profiles respectively, with the MED propagation model fixed.
One sees that the choice of DM halo profile has almost invisible impact on the determination of $X(E)$.
This is because the energetic leptons can not propagate a long distance and different DM profiles have very similar behavior except for the region near the Galaxy center.
The choice of propagation models do introduce large uncertainties into the determination of $X(E)$, but only for energies less than about $300$ GeV.
This is because, very high energy leptons must come from the neighborhood of the solar system.
It is reasonable to expect that the propagation effects should not have significant uncertainties in such a small distance.
Kinematically, the ICS gamma rays arising from the scattering on the CMB requires the initial $e^\pm$ energy $E_e \gsim m_e \sqrt{E_\gamma/\epsilon} /2$ ($\epsilon$ is the energy of CMB photon and $E_\gamma$ is the energy of the final ICS photon).
This means that the final ICS gamma rays with $E_\gamma \gsim 1$ GeV are produced from the initial electrons and positrons with $E_e \gsim 500$ GeV,
which has negligible uncertainties due to the choice of propagation model.
As a result, the predicted ICS gamma rays in the energy range of $1-10$ GeV have very small theoretical errors.
This implies that, adopting the conventional ``model 0'' background with the normalization factor $N=0.8$, decaying DM scenario fails to account for the $e^\pm$ excesses without violating gamma-ray upper limits of nearby clusters observed by the Fermi-LAT Collaboration.

\begin{figure}[tb]
\includegraphics[scale=1.0]{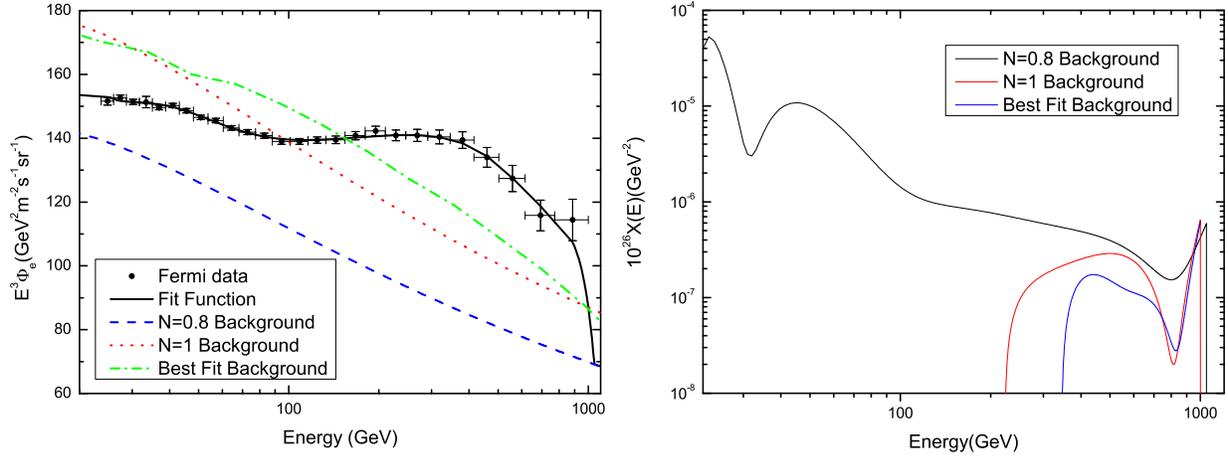}
\caption{\label{fig:more X} Left: Different astrophysical $e^{\pm}$ backgrounds as well as the Fermi-LAT data are shown. Right: Taking NFW halo profile and MED propagation model, $X(E)$ is determined from $f_{e}^\mathrm{DM}(E,r_{\odot})$ for various astrophysical $e^\pm$ backgrounds.}
\end{figure}

However our current understanding of the astrophysical $e^{\pm}$ backgrounds is still quite limited.
Recall that the normalization factor $\mathrm{N}=0.8$ in the conventional ``model 0'' background is chosen simply to leave room for the
additional DM component below 100 GeV. Actually, $\mathrm{N}=1$ was already used to interpret successfully low energy pre-Fermi data,
such as HEAT \cite{DuVernois:2001bb} and AMS-01 \cite{Aguilar:2002ad}. 
Recently, a full Bayesian analysis based on GALPROP was presented in \cite{Trotta:2010mx} to predict cosmic-rays self-consistently.
Taking their best fit parameters, the total $e^{\pm}$ background is found to be harder than the conventional ``model 0'' background. As shown in the left part of Fig. \ref{fig:more X}, the behavior of these background spectra below around $100$ GeV reveals potential inconsistency between the Fermi-LAT data and other observations at low energy. As a result, the DM induced $e^{\pm}$ spectra $f_{e}^{\mathrm{DM}}(E,\overrightarrow{r_{\odot}})$ at the earth, which can be obtained by subtracting off astrophysical $e^{\pm}$ background from the Fermi-LAT data, would even turn negative below $100$ GeV and $130$ GeV for $\mathrm{N}=1$ conventional background and the best fit background, respectively. This feature is certainly unphysical.

We thus focus on more energetic $e^{\pm}$. Eq. (\ref{eq:X}) tells us that the source spectrum $X(E)$ can be reconstructed from $f_{e}^{\mathrm{DM}}(E',\overrightarrow{r_{\odot}})$ at the earth with $E' \ge E$. This guarantees that the unphysical feature of $f_{e}^{\mathrm{DM}}(E,\overrightarrow{r_{\odot}})$ at low energy would not affect the determination of $X(E)$ at the higher end of the spectrum. 
Adopting MED propagation model and NFW DM halo model, we then reconstruct $X(E)$ via Eq. (\ref{eq:X}) for alternative choices of $e^{\pm}$ backgrounds, as plotted in the right part of Fig. \ref{fig:more X}. As stressed before, this determination of $X(E)$ is independent of 
any particle physics model of DM. Unsurprisingly, the inconsistency between the Fermi-LAT data and the $\mathrm{N}=1$ conventional $e^{\pm}$ background  (best fit $e^{\pm}$ background) at low energy leads to a negative source spectrum $X(E)$ below $220$ GeV ($340$ GeV) during our reconstruction procedure.
We do not show these unphysical spectra at low energy and, for simplicity, assume them to be vanishing in the right part of Fig. \ref{fig:more X}.
Fortunately, the GeV ICS photons are only sensitive to the initial $e^{\pm}$ with the energy $E_{e} \gtrsim 500$ GeV. 

\begin{figure}[tb]
\includegraphics[scale=1.0]{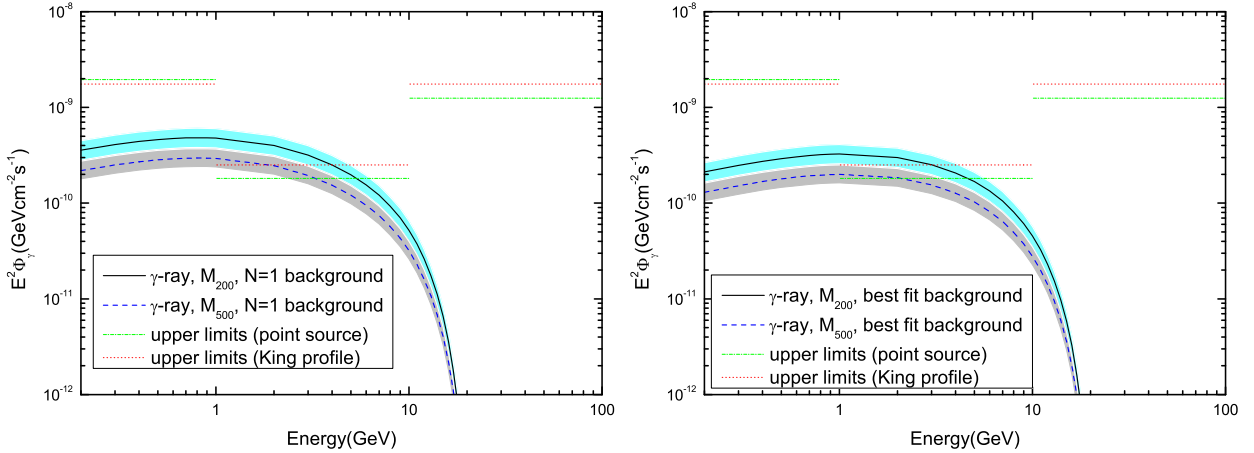}
\caption{\label{fig:more ICS}  The predicted ICS fluxes of photons in the Fornax cluster are shown in the left (right) part of the figure for the $\mathrm{N}=1$ conventional $e^\pm$ background (the best fit $e^{\pm}$ background). Experimental upper limits are taken from \cite{Ackermann:2010qj}.}
\end{figure}

Noticed that the alternative backgrounds lead to smaller fluxes of $f_{e}^{\mathrm{DM}}(E',\overrightarrow{r_{\odot}})$ in the whole energy range, compared to the $\mathrm{N}=0.8$ conventional background. As a result, the predicted ICS fluxes of photons should become softer.
Taking the Fornax cluster as a point source, we show in Fig \ref{fig:more ICS} the predicted ICS gamma rays in the Fornax cluster.
For the conventional ``model 0'' $e^{\pm}$ background with the normalization factor $\mathrm{N}=1$,
too much gamma-rays from the Fornax cluster are still predicted in the energy range 1\textendash{}10 GeV, which contradicts with the Fermi-LAT point-like upper limits. For the best fit $e^{\pm}$ background from a Bayesian analysis,
decaying DM scenario survives the experimental upper limits only if the Fornax cluster has a small total mass $M_{500}$.
This is a strong constraint since other processes besides ICS, such as the bremsstrahlung of energetic $e^{\pm}$ and $\pi^0$ decays, would also produce gamma rays. However these gamma rays can not be estimated in a model-independent way. Anyway, 
little room is left for these model-dependent gamma-ray fluxes from decaying DM in the Fornax cluster.

In summary, we have slightly improved the method to determine the DM-induced $e^\pm$ fluxes at the source from the corresponding fluxes at the earth, 
in a DM model independent way by solving the Volterra integral equation.
Accordingly, gamma rays emitted by these DM induced energetic leptons via ICS can be predicted in a model independent way. 
It is worth noticing that the DM-induced $e^\pm$ fluxes at the earth are obtained by subtracting off the astrophysical $e^\pm$ background from the Fermi-LAT measurements of the total flux of electrons and positrons.
So the prediction of ICS gamma rays depends on the choice of the astrophysical $e^\pm$ background, which is unfortunately not well determined.
As an illustration, we calculate the flux of ICS gamma rays from the Fornax cluster in the decaying DM scenario with different $e^\pm$ backgrounds.
For the conventional ``model 0'' $e^\pm$ background with the normalization factor $\mathrm{N}\leq1$,
the DM-induced ICS gamma rays from the Fornax cluster are found to exceed the upper limits measured by the Fermi-LAT Collaboration in the energy range of $1-10$ GeV. Using alternatively the best fit $e^{\pm}$ background from a Bayesian analysis, decaying DM scenario survives existing observations only if the Fornax cluster has a small total mass $M_{500}$. This is still a strong constraint as the gamma rays may be produced in other processes besides ICS, such as the bremsstrahlung and $\pi^0$ decays.
In addition, the DM-induced ICS gamma rays with $E_\gamma \gsim 1$ GeV are essentially independent of choices of propagation model and of DM density profile when a specific astrophysical $e^\pm$ background is assumed.

\textbf{Note added:} Two months after the first version of this paper appeared on arXiv, we noticed accidentally that the same kind of method had already  proposed in \cite{Hamaguchi:2009jb} to reconstruct the electron/positron source spectrum from the experimental fluxes at the earth. Comparing to \cite{Hamaguchi:2009jb}, we slightly improve this kind of method and  apply it to updated experimental data. Moreover, the ICS gamma rays from the Fornax cluster are also calculated in a DM-model independent way, which shows possible contradiction with or strong constraint from the Fermi-LAT measurements.   

\begin{acknowledgments}
This work is supported in part by the National Science Foundation
of China (No.10875103, No. 11075139 and No.10705024) and National
Basic Research Program of China (2010CB833000). M.L and G.Z are also
supported in part by the Fundamental Research Funds for the Central
Universities.
\end{acknowledgments}

\appendix

\section{The Volterra integral Equation}

The Volterra integral equation is given as
\begin{equation}
\int_{a}^{x}K(x-t)y(t)dt=f(x),
\end{equation}
with boundary condition $K(0)=1$ and $f(a)=0$.
The solution of $y(x)$ can be represented as
\begin{equation}
y(x)=\frac{df(x)}{dx}+\int_{a}^{x}dt\: R(x-t)\frac{df(t)}{dt}\label{eq:Solution of Volterra}
\end{equation}
with
\begin{equation}
R(x)=\mathrm{\boldsymbol{L}}^{-1}\left[\frac{1}{p\widetilde{K}(p)}-1\right] \mbox{~~and~~}
\widetilde{K}(p)=\mathrm{\boldsymbol{L}}\left[K(x)\right]~.
\end{equation}
Here $\mathrm{\boldsymbol{L}}$ denotes the Laplace transform and $\mathrm{\boldsymbol{L}}^{-1}$ its inverse. As a deformation of the Volterra integral equation, Eq. (\ref{eq:X}) can be easily obtained from Eq. (\ref{eq:Solution of Volterra}) by
a suitable change of variables.

 The kernel function in our case is $K(x)=\overset{{\scriptscriptstyle \infty}}{\underset{{\scriptscriptstyle {\scriptstyle m,n=1}}}{\sum}}B_{mn}\exp(\lambda_{mn}x)/\overset{{\scriptscriptstyle \infty}}{\underset{{\scriptscriptstyle {\scriptstyle m,n=1}}}{\sum}}B_{mn}$, correspondingly \begin{eqnarray}\label{eq:Cauchy}
 R(x)&=\mathrm{\boldsymbol{L}}^{-1}\left[-\overset{{\scriptscriptstyle \infty}}{\underset{{\scriptscriptstyle {\scriptstyle m,n=1}}}{\sum}}\frac{B_{mn}\lambda_{mn}}{p-\lambda_{mn}}/\overset{{\scriptscriptstyle \infty}}{\underset{{\scriptscriptstyle {\scriptstyle m,n=1}}}{\sum}}\frac{pB_{mn}}{p-\lambda_{mn}}\right] \nonumber \\
 &=\underset{i}{\sum}\,\mathrm{Res}\left[\frac{-\overset{{\scriptscriptstyle \infty}}{\underset{{\scriptscriptstyle {\scriptstyle m,n=1}}}{\sum}}\frac{B_{mn}\lambda_{mn}}{p_{i}-\lambda_{mn}}}{\overset{{\scriptscriptstyle \infty}}{\underset{{\scriptscriptstyle {\scriptstyle m,n=1}}}{\sum}}\frac{p_{i}B_{mn}}{p_{i}-\lambda_{mn}}}\exp(p_{i}x)\right]~,
\end{eqnarray}
where the Cauchy's Residue Theorem has been applied in the second line of the above equation, with $\mathrm{Res}$ denoting the residue. The summation $\underset{i}{\sum}$ is over all singularities $p_{i}$ in the left half complex plane. Defining $\psi(p)\equiv \overset{{\scriptscriptstyle \infty}}{\underset{{\scriptscriptstyle {\scriptstyle m,n=1}}}{\sum}}\frac{B_{mn}}{p-\lambda_{mn}}$, the singularities in Eq. (\ref{eq:Cauchy}) correspond to the zeros of $\psi(p)$ except $p=0$. In practice, the infinite summation in $\psi(p)$ should be truncated, in the same vein of Eq. (\ref{eq:Solution}). Then the number of zeros of $\psi(p)$ equals to the number of terms in the truncated series (e.g., as large as $1413$ for the case of MED propagation model and NFW DM density profile). We have checked numerically that the truncated terms with smaller $\lambda_{mn}$s have negligible effects on the position of relevant zeros. In principle, one can find all singularities and their residues in Eq. (\ref{eq:Cauchy}). But this demands excessive amount of computer power and it is unnecessary, as we will see presently. Notice that all singularities have negative real parts as $\lambda_{mn}<0$. Furthermore, there is an exponential suppression factor $\exp(p_{i}x)$ in the residue, which naturally offers a good damping factor. Therefore singularities in the region $-200 <\mathrm{Re}(p)<0$ should yield a very good approximation, as evidenced by the left part of Fig. \ref{fig:NFW MED}. Notice also that the terms being truncated in $\psi(p)$ contribute no additional singularity in the region $-200<\mathrm{Re}(p)<0$, as guaranteed by Rouche's theorem in complex analysis. So the errors of our method are well controlled.

To find the roots of $\psi(p)=0$ quickly, we apply the argument principle in complex analysis:
\begin{equation}
\frac{1}{2\pi i}\oint_\mathrm{C}\frac{\psi'(p)}{\psi(p)}\, dp=\mathrm{N}-\mathrm{P}, \label{eq:Argument principle}
\end{equation}
if $\psi(p)$ is a meromorphic function inside and on some closed
contour C and have no zeros or poles on C. N and P denotes the number
of zeros and poles of $\psi(p)$ inside the contour C respectively.
Each zero is counted as many times as its multiplicity while each
pole is counted as many times as its order. So we divide the region
$-200<\mathrm{Re}(p)<0$ into many small regions and do such a integral
around the contour of each small region to learn the distribution of the zeros of $\psi(p)$. Finally we use Newton's method to locate the zeros accurately in the corresponding small regions. This method is especially useful to locate the zeros off the real axis.

\end{document}